\input{aipcheck}

\documentclass[
  ,draft            % use draft while you are working on the paper
  ]
  {aipproc}

\layoutstyle{6x9}

\begin{document}

\title{XMM-Newton observations of XTE~J1817-330 and XTE~J1856+053}



\author{Gloria Sala}{
  address={Max-Planck-Institut f\"ur extraterrestrische Physik, 
  PO Box1312, 85741 Garching b.M., Germany}
}

\author{Jochen Greiner}{
  address={Max-Planck-Institut f\"ur extraterrestrische Physik, 
  PO Box1312, 85741 Garching b.M., Germany}
}

\author{Eugenio Bottacini}{
  address={Max-Planck-Institut f\"ur extraterrestrische Physik, 
  PO Box1312, 85741 Garching b.M., Germany}
}

\author{Frank Haberl}{
  address={Max-Planck-Institut f\"ur extraterrestrische Physik, 
  PO Box1312, 85741 Garching b.M., Germany}
}

\begin{abstract}
The black hole candidate XTE~J1817-330 was discovered in outburst on 26
January 2006 with RXTE/ASM.
One year later, on 28 February 2007, another
X-ray transient discovered in 1996, XTE~J1856+053,
was detected by RXTE during a new outburst. 
We report on the spectra obtained by XMM-Newton 
of these two black hole candidates.
\end{abstract}

\maketitle

%%%%%%%%%%%%%%%%%%%%%%%%%%%%%%%%%%%%%%%%%%%%
%% MAINMATTER
%%%%%%%%%%%%%%%%%%%%%%%%%%%%%%%%%%%%%%%%%%%%

%X-ray binaries are the brightest X-ray sources in the sky. They are powered
%by the accretion of material from the secondary star 
%onto a compact object (neutron star or black hole). 
%The generally accepted picture for the X-ray emission of accreting black holes consists of 
%an accretion disk, responsible for thermal black body emission in the X-ray band;
%and a surrounding hot corona, origin site of non-thermal power-law emission, 
%up to the energy range of gamma-ray telescopes, due to inverse comptonization 
%of soft X-ray photons from the disk.
%At present, around 20 X-ray binaries contain a dynamically confirmed black hole, and 
%around another 20 are the so called black-hole candidates \cite{rmc06}. 
%Seven of the 20 confirmed black holes, and 12 of the black-hole candidates
%are transient sources with only one unique outburst observed. 

\section{XTE~J1817-330}

\begin{figure}
\label{lcs}
  \includegraphics[height=.22\textheight]{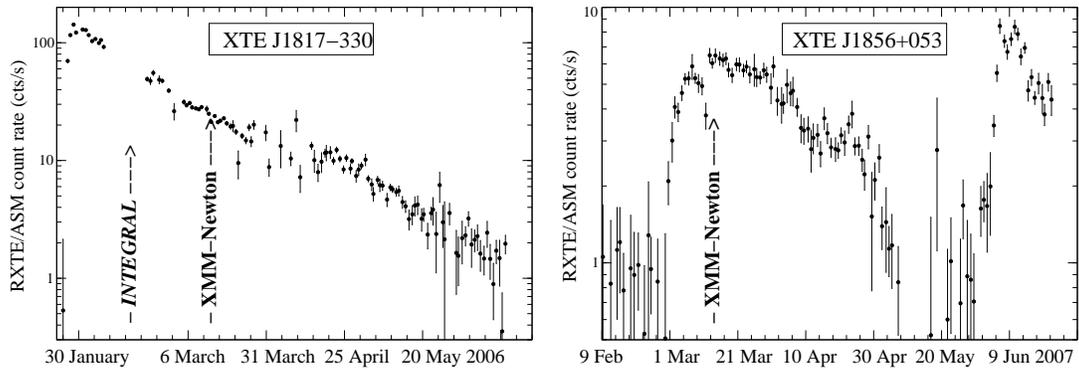}
  \caption{RXTE/ASM light-curves of XTE J1817-330 (left) and XTE J1856+053 (right). 
The dates of the XMM-Newton and INTEGRAL observations presented here are indicated.}
\end{figure}

XTE~J1817-330 was discovered by the Rossi X-ray Timing Experiment (RXTE) on
26 January 2006 \cite{rem06} with a flux of 0.93($\pm0.03$)~Crab (2-12 keV) and 
a very soft spectrum, typical for black hole transients.
We obtained a Target of Opportunity Observation (TOO) 
with XMM-Newton (0.1-10.0~keV) on 13 March 2006 (obs. ID. 0311590501, 20~ks), 
when the source flux detected by 
the All Sky Monitor (ASM) on board RXTE had faded to $\sim$300~mCrab \cite{sal06}
(left panel in Fig.~\ref{lcs}).

We fit simultaneously the data of all XMM-Newton instruments 
active during the observation:
EPIC-pn (0.6--10.0~keV; Burst mode), RGS1 (0.3--2.0~keV) and 
OM (U and UWV1 filters) (Fig.~\ref{fig}).
We use a 2-component model consisting of a thermal accretion disk 
({\it diskpn} model available in {\it xspec} \cite{gie99})
plus a comptonization component ({\it compTT} model \cite{tit94}) 
to fit the spectrum of XTE~J1817-330. 
The inner radius in the {\it diskpn} model is fixed to 6$R_g$.
We obtain the best fit for the {\it diskpn + compTT} model ($\chi^2_{\nu}=1.18$) with 
$N_{\rm{H}}=1.55(\pm0.05)\times10^{21}\,\mbox{cm}^{-2}$,
a disk with $kT_{\rm{in}}=0.70(\pm0.01)\,\mbox{keV}$ and a
comptonization component with kT$_{e}=50$~keV (fixed) and $\tau=0.15(\pm0.02)$.
The observed X-ray flux is 
$8.6(\pm 0.8)\times10^{-9}\,\mbox{erg}\,\mbox{cm}^{-2}\,\mbox{s}^{-1}$ (0.4-10~keV), 
and the unabsorbed X-ray luminosity of the source at the time of the observation 
L$_{(0.4-10\,keV)}=1.2(\pm0.1)\times10^{38}({D}/10\rm{kpc})^2\,\mbox{erg}\,\mbox{s}^{-1}$.

The normalization constant {\it K} of the {\it diskpn} model is 
related to the mass of the compact object {\it M}, the 
distance to the source {\it D}, and the inclination {\it i} of the disk as 
$K=\frac{M^2 cos(i)}{D^2 \beta^4}$,
where $\beta$ is the color/effective temperature ratio. 
Furthermore, the accretion rate can be obtained
from the mass of the compact object and the maximum temperature of the disk \cite{gie99}. 
Assuming $\beta=1.7$ and using the best fit value for the normalization of the 
{\it diskpn} model ($K_{\rm {diskpn}}=0.024\pm0.002$), we can compare the accretion rate 
for different possible masses, distances, and inclinations with an
upper limit for the accretion rate. 
At the time of the XMM-Newton observations, the flux of the source 
had decreased by a factor 6 with respect to the maximum registered by RXTE. 
Taking the Eddington limit as the upper limit for 
the accretion rate at the maximum of the burst,
the accretion rate at the time of the observation could not be higher than 
16\% of $\rm M^{\rm{acc}}_{\rm{Edd}}$.
This sets an upper limit for the mass of the central object of 6~M$_{\odot}$
(see \cite{salsub} for more details).

INTEGRAL observed XTE~J1817-330 in hard X-rays (20-150~keV) 
as a TOO on 15-18 February 2006 \cite{gol06} (200~ks exposure). 
We fit simultaneously JEM-X (6--30~keV) and IBIS/ISGRI (20--200~keV) spectral data 
obtained on 15--18 February 2006 with a two component model, \emph{diskbb + pow}.
%We add a free relative normalization constant to IBIS data 
%to account for differences between the JEM-X and IBIS calibration. 
The best fit ($\chi^2_\nu=1.3$) is obtained with $kT_{\rm{in}}=0.95(\pm0.04)\,\mbox{keV}$, 
$K_{\rm {diskbb}}=\left( \frac{R_{in}/km}{D/10kpc}\right)^2 \cos i=1500\pm400$, 
photon index $\Gamma=2.64\pm0.04$ and power-law normalization 
$K=4.2\pm0.4\,\mbox{ph}\,\mbox{keV}^{-1}\,\mbox{cm}^{-2}\,\mbox{s}^{-1}$@1~keV.

\section{XTE~J1856+053}

On 28 February 2007 a new outburst of the previously known transient source 
XTE~J1856+053 was detected with RXTE \cite{lev07}.
We obtained a TOO observation with XMM-Newton (0.1-10.0~keV) 
on 14 March 2007 (obs. ID. 0510010101, 5~ks for RGS, 1.5~ks for EPIC-pn), 
when the source flux detected by the RXTE/ASM was maximum, $\sim$80~mCrab \cite{sal07}
(right panel in Fig.~\ref{lcs}).

\begin{figure}
\label{fig}
  \includegraphics[height=.26\textheight]{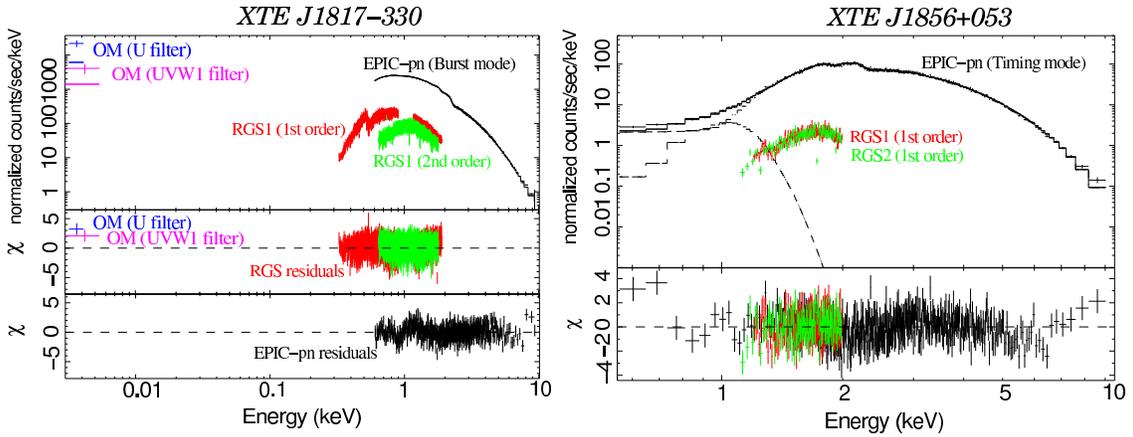}
  \caption{XMM-Newton spectra of XTE J1817-330 (left) and XTE J1856+053 (right).}
\end{figure}

We fit simultaneously XMM-Newton EPIC-pn (0.55--10.0~keV, in Timing mode), 
RGS1 and RGS2 (0.3--2.0~keV) data (Fig.~\ref{fig}).
We fit the spectrum with a thermal accretion disk 
({\it diskpn}). 
No indication of a hard component is evident in the residuals, 
but an excess is present below 1 keV, leading to a poor reduced 
$\chi^2$ of 1.99. Adding a recombination emission edge
at 0.87~keV (corresponding to O~VIII~K-shell) with
plasma temperature kT$=50(\pm3)$~eV improves the fit.
The significance of this feature is however to be taken with 
care, since the excess could be caused by some redistribution of 
higher energy photons to lower energies not properly 
taken into account by the calibration.
The best fit ($\chi^2_{\nu}=1.16$) is then obtained with 
$N_{\rm{H}}=4.45(\pm0.05)\times10^{22}\,\mbox{cm}^{-2}$, and
a disk with $kT_{\rm{in}}=0.76(\pm0.01)\,\mbox{keV}$ and 
$K_{\rm {diskpn}}=\frac{M^2 \cos i}{D^2 \beta^4}=(8.5\pm0.4)\times10^{-3}$.
The observed X-ray flux is 
$1.0(\pm0.1)\times10^{-9}\,\mbox{erg}\,\mbox{cm}^{-2}\,\mbox{s}^{-1}$ 
(0.5--10.0~keV), which corrected for absorption corresponds to an unabsorbed X-ray luminosity 
L$_{(0.5-10.0\,keV)}=4.0(\pm1.5)\times10^{38}({D}/10\rm{kpc})^2\,\mbox{erg}\,\mbox{s}^{-1}$.

The low temperature of the accretion disk favours a black-hole as the 
accreting compact object. 
With the normalization constant {\it K} of the {\it diskpn} model 
we can put some contraint to the mass of the compact object {\it M}, 
as done above for XTE J1817-330.
At the time of the XMM-Newton observations, the flux of the source 
was at the maximum of the first burst detected in March 2007. However, a brigther
burst was detected by RXTE/ASM in June 2007. 
At the time of our XMM-Newton observations, the flux was 70\% 
of the maximum detected in June 2007. 
Taking the Eddington limit as the upper limit for the accretion rate at the 
maximum in June 2007, the accretion rate at the time of the XMM-Newton 
observation could not be higher than  $0.7\rm M^{\rm{acc}}_{\rm{Edd}}$.
This sets an upper limit for the mass of the central object of 4.2~M$_{\odot}$. 

%%%%%%%%%%%%%%%%%%%%%%%%%%%%%%%%%%%%%%%%%%%%%%%%
%% BACKMATTER
%%%%%%%%%%%%%%%%%%%%%%%%%%%%%%%%%%%%%%%%%%%%%%%%

\begin{theacknowledgments}
We thank Norbert Schartel and the XMM-Newton team for carrying out the 
TOO observations presented here. 
XMM-Newton and INTEGRAL projects are ESA Science Missions
directly funded by ESA Member States and the USA (NASA),
with support from BMWI/DLR (FKZ 50 OX 0001), the Max-Planck
Society and the Heidenhain-Stiftung.
GS and EB are supported through DLR (FKZ 50 OR 0405).
\end{theacknowledgments}



\bibliographystyle{aipprocl} % if natbib is missing

%\bibliography{sample}

%  \typeout{******************************************}
%  \typeout{** Please run "bibtex \jobname" to optain}
%  \typeout{** the bibliography and then re-run LaTeX}
%  \typeout{** twice to fix the references!}
%  \typeout{******************************************}

%%%%%%%%%%%%%%%%%%%%%%%%%%%%%%%%%%%%%%%%%%%
%% The following lines show an example how to produce a bibliography
%% without the help of the BibTeX program. This could be used instead
%% of the above.
%%%%%%%%%%%%%%%%%%%%%%%%%%%%%%%%%%%%%%%%%%%

%\endinput
%%
%% End of file `template-6s.tex'.
\end{document}